\documentclass{IEEEtran4PSCC}

\usepackage[pdftex]{graphicx}
\usepackage[cmex10]{amsmath}
\usepackage{gensymb}
\usepackage{xcolor}

\usepackage{float}

\begin{document}

\newcommand{\indexnode}{i}
\newcommand{\indexconductor}{p}
\newcommand{\indextime}{k}
\newcommand{\indexconverter}{c}
\newcommand{\indexstorage}{s}

\newcommand{\setconductor}{\mathcal{P}}
\newcommand{\setTime}{\mathcal{K}}

\newcommand{\symtime}{T}
\newcommand{\symenergy}{E}
\newcommand{\symcharge}{Q}

\newcommand{\symcomplexpower}{S}
\newcommand{\symactivepower}{P}
\newcommand{\symreactivepower}{Q}
\newcommand{\symcurrent}{I}
\newcommand{\symcurrentlifted}{L}
\newcommand{\symvoltage}{U}
\newcommand{\symvoltagelifted}{W}
\newcommand{\symimpedance}{Z}
\newcommand{\symresistance}{R}
\newcommand{\symreactance}{X}
\newcommand{\symstatus}{s}
\newcommand{\symbinary}{z}
\newcommand{\symefficiency}{\eta}

\newcommand{\paramcolor}{black}

\newcommand{\imag}{j}

\newcommand{\Scrating}{\textcolor{\paramcolor}{\symcomplexpower_{\indexconverter}^{\text{rating}}}}
\newcommand{\Icrating}{\textcolor{\paramcolor}{\symcurrent_{\indexconverter}^{\text{rating}}}}

\newcommand{\Scratingp}{\textcolor{\paramcolor}{\symcomplexpower_{\indexconverter, \indexconductor}^{\text{rating}}}}
\newcommand{\Icratingp}{\textcolor{\paramcolor}{\symcurrent_{\indexconverter, \indexconductor}^{\text{rating}}}}

\newcommand{\statusct}{\textcolor{\paramcolor}{\symstatus_{\indexconverter, \indextime}}}

\newcommand{\Sct}{\symcomplexpower_{\indexconverter, \indextime}}
\newcommand{\Pct}{\symactivepower_{\indexconverter, \indextime}}
\newcommand{\Qct}{\symreactivepower_{\indexconverter, \indextime}}

\newcommand{\Scpt}{\symcomplexpower_{\indexconverter, \indexconductor, \indextime}}
\newcommand{\Pcpt}{\symactivepower_{\indexconverter, \indexconductor, \indextime}}
\newcommand{\Qcpt}{\symreactivepower_{\indexconverter, \indexconductor, \indextime}}

\newcommand{\Pcat}{\symactivepower_{\indexconverter, a, \indextime}}
\newcommand{\Pcbt}{\symactivepower_{\indexconverter, b, \indextime}}
\newcommand{\Pcct}{\symactivepower_{\indexconverter, c, \indextime}}

\newcommand{\Qcat}{\symreactivepower_{\indexconverter, a, \indextime}}
\newcommand{\Qcbt}{\symreactivepower_{\indexconverter, b, \indextime}}
\newcommand{\Qcct}{\symreactivepower_{\indexconverter, c, \indextime}}

\newcommand{\Ict}{\symcurrent_{\indexconverter, \indextime}}
\newcommand{\Lct}{\symcurrentlifted_{\indexconverter, \indextime}}

\newcommand{\Icpt}{\symcurrent_{\indexconverter,  \indexconductor,\indextime}}
\newcommand{\Lcpt}{\symcurrentlifted_{\indexconverter, \indexconductor, \indextime}}

\newcommand{\Uit}{\symvoltage_{\indexnode, \indextime}}
\newcommand{\Zc}{\textcolor{\paramcolor}{\symimpedance_{\indexconverter}}}

\newcommand{\Wit}{\symvoltagelifted_{\indexnode, \indextime}}

\newcommand{\Uipt}{\symvoltage_{\indexnode, \indexconductor, \indextime}}
\newcommand{\Zcp}{\textcolor{\paramcolor}{\symimpedance_{\indexconverter, \indexconductor}}}
\newcommand{\Rcp}{\textcolor{\paramcolor}{\symresistance_{\indexconverter, \indexconductor}}}
\newcommand{\Xcp}{\textcolor{\paramcolor}{\symreactance_{\indexconverter, \indexconductor}}}

\newcommand{\Uiptmin}{\textcolor{\paramcolor}{\symvoltage_{\indexnode, \indexconductor}^{\text{min}}}}
\newcommand{\Uiptmax}{\textcolor{\paramcolor}{\symvoltage_{\indexnode, \indexconductor}^{\text{max}}}}

\newcommand{\Wipt}{\symvoltagelifted_{\indexnode, \indexconductor, \indextime}}

\newcommand{\Pstor}{\symactivepower_{\indexconverter, \indextime}^{\text{stor}}}
\newcommand{\Sstor}{\symcomplexpower_{\indexconverter, \indextime}^{\text{stor}}}

\newcommand{\Pext}{\textcolor{\paramcolor}{\symactivepower_{\indexconverter, \indextime}^{\text{ext}}}}
\newcommand{\Sext}{\textcolor{\paramcolor}{\symcomplexpower_{\indexconverter, \indextime}^{\text{ext}}}}

\newcommand{\Qint}{\textcolor{\paramcolor}{\symreactivepower_{\indexconverter, \indextime}^{\text{int}}}}

\newcommand{\Pcstor}{\symactivepower_{\indexconverter, \indextime}^{\text{c}}}
\newcommand{\Pdstor}{\symactivepower_{\indexconverter, \indextime}^{\text{d}}}

\newcommand{\bc}{\symbinary_{\indexconverter, \indextime}^{\text{c}}}

\newcommand{\Erating}{\textcolor{\paramcolor}{\symenergy_{\indexconverter}^{\text{max}}}}
\newcommand{\Pcstorrating}{\textcolor{\paramcolor}{\symactivepower_{\indexconverter}^{\text{c,max}}}}
\newcommand{\Pdstorrating}{\textcolor{\paramcolor}{\symactivepower_{\indexconverter}^{\text{d,max}}}}

\newcommand{\etacstor}{\textcolor{\paramcolor}{\symefficiency_{\indexconverter}^{\text{c}}}}
\newcommand{\etadstor}{\textcolor{\paramcolor}{\symefficiency_{\indexconverter}^{\text{d}}}}

\newcommand{\Ect}{\symenergy_{\indexconverter, \indextime}}
\newcommand{\Ectinit}{\textcolor{\paramcolor}{\symenergy_{\indexconverter}^{\text{init}}}}
\newcommand{\Ectprev}{\symenergy_{\indexconverter, \indextime-1}}

\newcommand{\Ectone}{\symenergy_{\indexconverter, \indextime=1}}

\newcommand{\Ectend}{\symenergy_{\indexconverter, \indextime=n}}

\newcommand{\Tt}{\textcolor{\paramcolor}{\symtime_{\indextime}}}

\title{A Flexible Storage Model for Power Network Optimization}

\author{\IEEEauthorblockN{Frederik Geth\IEEEauthorrefmark{1},
Carleton Coffrin\IEEEauthorrefmark{2},
David M Fobes\IEEEauthorrefmark{2}}
\IEEEauthorblockA{\IEEEauthorrefmark{1} Commonwealth Scientific and Industrial Research Organisation (CSIRO)\\
Canberra, Australia\\
frederik.geth@csiro.au}
\IEEEauthorblockA{\IEEEauthorrefmark{2} Los Alamos National Laboratory (LANL)\\
Los Alamos, New Mexico, USA\\
\{dfobes,cjc\}@lanl.gov}
}

\maketitle

\begin{abstract}
This paper proposes a simple and flexible storage model for use in a variety of multi-period optimal power flow problems. The proposed model is designed for research use in a broad assortment of contexts enabled by the following key features:  (i) the model can represent the dynamics of an energy buffer at a wide range of scales, from residential battery storage to grid-scale pumped hydro; (ii) it is compatible with both balanced and unbalanced formulations of the power flow equations; (iii)  convex relaxations and linear approximations to allow seamless integration of the proposed model into applications where convexity or linearity is required are developed; (iv) a minimalist and standardized data model is presented, to facilitate easy of use by the research community.  The proposed model is validated using a proof-of-concept twenty-four hour storage scheduling task that demonstrates the value of the model's key features.  An open-source implementation of the model is provided as part of the PowerModels and PowerModelsDistribution optimization toolboxes.
\end{abstract}


\section{Introduction}

As the penetration of renewable energy resources and electric vehicles on modern power grids continues to increase, so does the amount of uncertainty in the network's power balance.  Energy storage presents a natural solution to mitigating these uncertainties by smoothing short-term fluctuations in load and generation.  However, modeling energy storage devices presents three fundamental challenges: (1) a variety storage technologies exists, such as pumped hydro, compressed air, flywheels, molten salt, and batteries, each with unique characteristics; (2) many storage devices can provide valuable ancillary services, \textit{e.g.}, frequency control and reactive power support, which are important to capture when making operational or investment decisions; (3) storage devices are available in an assortment of sizes, from massive bulk electric system facilities to small household batteries.  The combination of these features presents a notable challenge to modeling storage devices in the context of network operations and investment.

Inspired by the success of the $\Pi$-equivalent circuit for modeling branches in transmissions systems, this work proposes a generic and flexible grid-connected storage model that strives to strike a balance between model simplicity and flexibility.

In the most general case, the proposed storage model captures a three-phase steady-state AC grid connection with a variety of losses, suitable for detailed operational models in power distribution systems.  Consistent single-phase equivalent and active-power-only approximations of this model, which are suitable for applications requiring less detail, such as transmission network investment planning have been developed with care.  These mathematical variants provide a natural trade-off between model accuracy and scalability.  Furthermore, convex relaxations of the storage model are proposed to provide lower bounds on AC Optimal Power Flow (OPF) studies with storage devices.

This paper begins with a review and analysis of storage simulation and optimization models for power system operation in Section \S \ref{sec_lit_review}.
A research-grade, easy-to-use, simplified storage model is then developed in Section \S \ref{sec_model}.
Relaxations and approximations of the nonlinear model are proposed in Section \S \ref{sec_model_relax}.
Section \S \ref{sec_study} illustrates the use of the proposed models in the context of reducing generation fuel costs in power grid operations, and Section \S \ref{sec_conclusions} concludes the paper.

\section{Literature Review}\label{sec_lit_review}
Both \cite{Das2018} and \cite{Diaz-Gonzalez2012} provide detailed discussions of the storage technology options and have focus on how to best support renewable integration in power systems.
In contrast, this section provides a high level overview of storage models with a focus on optimal operation.

\paragraph*{Energy Storage}
Storage technologies can be broadly categorized by the nature of their energy conversion processes:  batteries (\textit{e.g.}, in electric vehicles or stationary storage) rely on a reversible electrochemical process,  pumped hydro on a gravitational field,  flywheels on a kinetic energy system, and pressure is leveraged by compressed air storage.  Interestingly, grid-level energy storage need-not be bi-directional.  One such case is thermal storage, where electric power is converted to heat preemptively to be utilized later (\textit{e.g.}, in residential hot water heaters).  Another common case is demand response protocols, which can modulate power consumption temporarily to time-shift power demands.
In this work, these uni-directionally interfaced systems are treated as a special case of a bi-directional model.

Consuming power from the grid generally means the energy buffer increases, which is referred to as storing energy.  Injecting power into the grid results in reductions in the energy buffer.
In the context of battery storage, we use the terminology of charging (energy content increases) and discharging  (energy content decreases)  the batteries; for pumped hydro it is often called pumping and turbining.
In both cases, the underlying principle for storing energy is through a buffer $E$,
\begin{IEEEeqnarray}{C}
P(t) = \frac{dE(t)}{dt}
\end{IEEEeqnarray}
where a nonzero power $P(t)$ will lead to a change in energy $E(t)$ over time $t$.

Regardless of storage approach, the key feature that distinguish storage devices from other grid connected devices is that their energy content can be controlled independently of the electrical interface, and can therefore be optimized to meet objectives like cost reduction costs or quality-of-service increases.  However, it is important to note that energy losses occur during charging, discharging, and while idle (\textit{e.g.}, energy dissipation).  Such losses are important contributors to the overall costs of operating and optimizing such systems.

\paragraph*{Storage System Simulation}
For simulation it is common to develop highly fidelity storage device models.  For instance, \cite{Tremblay2007} developed a detailed electrical battery storage model for validation of electric vehicle simulation, which incorporates the dependence of battery voltage on the state of charge, \textit{i.e.},
\begin{IEEEeqnarray}{C}
U(t) = f(Q(t))
\end{IEEEeqnarray}
and
 models a charge buffer $Q$ instead of an energy buffer, \textit{i.e.},
\begin{IEEEeqnarray}{C}
I(t) = \frac{dQ(t)}{dt}
\end{IEEEeqnarray}
It is noted that the charge-dependency of the terminal voltage is very limited with technologies such as Li-ion over a broad range of state of charge, therefore these effects are commonly ignored.
Furthermore, when the battery voltage is assumed to be constant, \textit{e.g.}, $U^{\text{nom}}$, in the normal operation range, one can represent the charge buffer $Q$ as an energy buffer $E$,
\begin{IEEEeqnarray}{C}
P(t) \approx U^{\text{nom}} \cdot I(t) = U^{\text{nom}} \cdot \frac{dQ(t)}{dt} =  \frac{dE(t)}{dt},
\end{IEEEeqnarray}
where a nonzero power $P(t)$ will lead to a change in energy $E(t)$ over time $t$.
Such an energy buffer expression can easily represent other storage processes.
For simulation, these equations are solved using numerical integration methods.

High fidelity models for simulation of storage systems are available in such tools as OpenDSS \cite{Dugan2011,Dugan2013a} and GridLAB-D \cite{gridlabd}.  However, these simulators rely on local control heuristics or user-provided schedules for storage device operation decisions.  A natural solution would be embed these models into mathematical {\em optimization} tool chains, where it is possible to define an objective, and allow an algorithm to determine the ideal operation of the storage system.  Unfortunately, these detailed storage models are often too complex for state-of-the-art optimization tools and hence there is a need to balance the storage model's accuracy for the abilities of modern optimization tools.

To that end, a number of papers have proposed optimization-compatible storage models, including \cite{Tant12,Eyisi2019,Alnaser2015,hari2018hierarchical} and similar models have been incorporated into software tools such as {\sc Matpower}'s MOST \cite{Zimmerman2011,MOST19} and pandapower \cite{Meinecke2018}.
We note that Heussen et al.  \cite{Heussen2011} developed a approach for representing different storage technologies through an optimization-compatible parameterized mathematical abstraction called the `Power Nodes Framework'.
These models have varying levels of sophistication, however we find that their scope is relatively narrow in terms of technologies \emph{and} support for different power flow representations.  Specifically, they exclusively focus on single-phase models with an active-power emphasis; it is not immediately clear how these models could be generalized to consider all of the modeling contexts targeted by this work.

\paragraph*{Unit Commitment and Multi-Period OPF}
Pumped hydro has historically been the most popular electricity storage technology (bidirectionally interfaced). Therefore, substantial work has occurred in the context of economic dispatch and unit commitment, including pumped hydro units \cite{Almassalkhi2016,Bruninx2016a}. Typically, such approaches have very limited representations of the grid's electrical physics; for instance, the `DC', network flow, and copper plate approximations are commonly used.

Nevertheless, driven by the interest in storage in distribution grids, extending optimal power flow with storage models has been a topic of interest \cite{Jabr2015}. Commonly, optimal power flow is considered as a snap-shot problem, \textit{i.e.}, all time steps are solved independently. Extending OPF to include multiple time steps simultaneously is commonly referred to as multi-period OPF \cite{Electric2013,Gemine2014b}. In multi-period OPF frameworks, one can incorporate battery storage models, including representations of the energy storage dynamics and/or other intertemporal constraints, such as ramp rates. Kouronis et al. \cite{Kourounis2018} demonstrate that large-scale multi-period AC OPF with storage (no complementarity constraint) can be done using customized interior-point NLP codes. Stai et al. propose a successive convex approximation scheme to improve scalability \cite{Stai2018}.
One can further extend such frameworks with higher fidelity physical models to represent phase unbalance \cite{Connell2014, Stai2018}.

\paragraph*{Balanced and Unbalanced Optimal Power Flow}
Optimal power flow has historically focused on modelling transmission networks, where assuming that the grid impedance and the loading allow for balanced voltage and current phasors is typically considered adequate.
In distribution systems, and especially in low-voltage grids, these assumptions are difficult to justify.
In this context, unbalanced OPF has been proposed as a generalization of OPF where phase unbalance effects are explicitly modeled.

Storage devices are particularly interesting in unbalanced OPF, as one can choose separate power set-points for each phase of the converter, assuming the technology allows it. In the most general case, both active and reactive power set-points can be chosen independently for each phase, subject to the active power being balanced with the energy storage subsystem while charging or discharging. Nevertheless, even without a storage subsystem, converters can circulate active power between phases. Such functionality can be delivered by inverter-based systems \cite{Stuyts2018}. The differences in control freedom have been explored \cite{Tant2015a}, but software implementations for modeling such features remained unavailable.

\paragraph*{Convex Relaxation}
Recently, for both balanced and unbalanced OPF, a significant amount of work has been published on reformulating and relaxing the mathematical models to obtain better numerical performance and stronger optimality guarantees \cite{Molzahn2017a}.
One of the key works describing relaxations of unbalanced OPF is \cite{Gan2014}.
To the best of our knowledge, few works have considered the use of storage optimization models in combination with convex relaxation of the power flow physics.  Papers that do derive such storage models include \cite{Marley2016,Geth2016}.

Careful consideration must be taken when combining published optimization formulations for storage systems with OPF formulations.
Intersecting two preexisting feasible sets with known complexity may require the definition of new constraints to serve as bridges between the variable spaces, potentially increasing the overall complexity.
For instance, a current-voltage variable space with a linear formulation of the branch physics, when combined with a linear power-voltage model for a generator, requires the addition of the definition of complex power, $S = U(I)^{*}$, as a bridge; this equation is nonlinear, which makes the integrated problem nonlinear (and non-convex).
To mitigate this, this paper defines a number of storage models with different complexities, and in a number of variable spaces that are compatible with recently developed formulations for balanced and unbalanced OPF.

\paragraph*{Discussion}
In summary, we observe that there is a need to develop a research-grade mathematical model for representing a broad range of storage technologies, which is suitable for different levels of grid modeling fidelity, \textit{e.g.}, AC NLP vs linearized `DC' and single-phase vs multi-phase, and is easy to implement in various grid optimization applications.  Developing such a model is the objective of this work, and a detailed derivation is provided in the subsequent sections.
Next to the features that it offers, we also want it to be convenient template to implement extensions for, e.g., for storage sizing, of technology-specific detailed models such as for battery storage systems.

\section{The Proposed Storage Model} \label{sec_model}
In this section we develop a foundational mathematical model for a simple energy storage system composed of two core parts: (1) a flexible energy storage subcomponent, which supports adaptive time discretization; and (2) a grid interface that can represent both inverters and synchronous machines.  The derivation begins with a general grid interface, which is a nonlinear multi-conductor model.  It then integrates that interface into the energy buffer.

\subsection{Time Discretization}
Using differential equations directly in optimization models is often prohibitive.
Consequently the first step in developing this model is to discretize time.
Natural time is defined by $t$. Time is sampled at instances $t \in \mathcal{T}$, and there exists a bijection with respect to indices $k  \in \mathcal{K}$ (integers). The time step length $\Tt$ is the amount of time between time $t_k$ and $t_{k+1}$. Note that we do not assume this to be uniform.
Therefore, continuous time symbols map to discrete time as follows:
\begin{IEEEeqnarray}{C}
P(t) \rightarrow P_k,
E(t) \rightarrow E_k
\end{IEEEeqnarray}
Next, we integrate the ODE w.r.t. the sample times, to obtain approximate algebraic expressions.
\begin{IEEEeqnarray}{C}
\int_{t_k}^{t_k+\Tt} P(t)dt = \int_{t_k}^{t_k+\Tt} dE(t) = E_{\indextime+1} -  E_{\indextime} \label{eq_int_e}
\end{IEEEeqnarray}
Integrating the $P(t)dt$ term approximately via the trapezoidal rule yields,
\begin{IEEEeqnarray}{C}
\int_{t_k}^{t_k+\Tt} P(t)dt \approx \left(\frac{P_{\indextime+1} +  P_{\indextime}}{2}\right) \Tt \label{eq_trap}
\end{IEEEeqnarray}
With the following special cases for in the initial and final points respectively,
\begin{IEEEeqnarray}{C}
\int_{t_k}^{t_k+\Tt} P(t)dt \approx P_{\indextime} \Tt \label{eq_init} \\
\int_{t_k}^{t_k+\Tt} P(t)dt \approx P_{\indextime+1} \Tt  \label{eq_final}
\end{IEEEeqnarray}
Note that \eqref{eq_trap} is exact when $P(t)$ is piece-wise linear with breakpoints at $t_k$ and $t_k+\Tt$, and \eqref{eq_init},\eqref{eq_final} are exact when $P(t)$ is a step function (piece-wise constant) with break points at $t_k$ and $t_k+\Tt$.
Combining \eqref{eq_int_e} together with any of \eqref{eq_trap},\eqref{eq_init},\eqref{eq_final} results in an algebraic expression, which can be generated for any sampled time $t_k$. This results in a set of \emph{sparse} difference equations that link the power and energy variables.

Note that in all cases, $t$ and $\Tt$ are known, but we want to decide on the values of $P_{\indextime}\,\,\forall \indextime$, which then determines $E_{\indextime}$. Both $E_{\indextime}$ and $P_{\indextime}$ must nevertheless remain within user-specified ranges. So, it is not a valid decision discharge when the energy content is at its lower bound (empty), or charge when the energy content is at the upper bound (full). The fact that a decision now, \textit{e.g.}, to charge at a certain power setting, impacts the future decision making freedom, suggests that an optimal decision may not be obvious.

Leveraging this derivation, the proposed optimization model is then defined in descretized time with time steps $\indextime \in \setTime = \{1,2,\dots, n \}$ and $\Tt$ being the duration of time step $\indextime$.  Note that time step durations need not be uniform.

\subsection{Nonlinear Multi-Conductor Converter Model}
The complex power flow from the grid into the converter $\indexconverter$, through conductor $\indexconductor$, at time $\indextime$, is given by,
\begin{IEEEeqnarray}{C}
\Scpt = \Pcpt + \imag \Qcpt. \label{eq_power}
\end{IEEEeqnarray}
We define the set of conductors (phases), $\indexconductor \in \setconductor$.
This power flow is subject to a per-conductor apparent power flow limit $\Scratingp$ and a time-dependent operational status 0/1 parameter $\statusct$,
\begin{IEEEeqnarray}{C}
  |\Scpt|^2 =   (\Pcpt)^2 + (\Qcpt)^2 \leq (\statusct \cdot \Scratingp)^2
    \end{IEEEeqnarray}
Current limits are defined per conductor,
\begin{IEEEeqnarray}{C}
  |\Icpt|  \leq \statusct \cdot \Icratingp
      \end{IEEEeqnarray}
  The relationship between power, current and voltage magnitudes is given by,
\begin{IEEEeqnarray}{C}
     |\Scpt|^2 = |\Uipt|^2 |\Icpt|^2. \label{eq_complex_power_def}
     \end{IEEEeqnarray}
Combining these properties, the converter's power balance and losses are defined as,
\begin{IEEEeqnarray}{C}
 \!\!\!\!\!\!\underbrace{ \sum_{\indexconductor \in \setconductor}  \Scpt }_{\text{grid side power}} \!+ \!\!\!\!\!\!\underbrace{\Pstor}_{\text{energy buffer}} \!\!\! =  \!\!\! \underbrace{\imag \Qint}_{\text{reactive source}} \!\!\! + \!\! \underbrace{\Sext}_{\text{other flow}} \!\!\!+\!\!  \underbrace{\sum_{\indexconductor \in \setconductor} \Zcp |\Icpt|^2}_{\text{copper loss}} \label{eq_balance_loss}
\end{IEEEeqnarray}
This equation is composed of five components (left to right):
\begin{itemize}
    \item the complex power flow supplied through the all of the converter's conductors $\Scpt$;
    \item the storage subsystem active power $\Pstor$;
    \item a variable $\Qint$ that represents the converter's ability to control the generation and/or absorption of reactive power;
    \item the user-defined complex power $\Sext$;
    \item copper losses proportional to internal\footnote{effectively modelling the converter as a voltage source with a series impedance} impedance $\Zcp$ and current magnitude $\Icpt$ squared.
\end{itemize}
Note that the inclusion of a copper loss term, due to its quadratic nature, incentivizes charging and discharging slowly over time, as well as avoiding unnecessary reactive power injection or consumption.  This has an added benefit of avoiding degeneracy issues related to multiple reactive power sources and sinks on a single node.  If reactive power is absorbed or generated in the inverter, the reactive power in \eqref{eq_balance_loss} remains balanced through variable $\Qint$.

The $\Sext$ parameter often represents unavoidable losses that are incurred while the system is idle, but it can also be used to capture a wide variety of exogenous energy sinks or sources. Examples of exogenous flows are, self-discharge in a battery,  heat losses from a heat  buffer to the environment, water flow into the upper reservoir of a pumped hydro plant.

\subsection{Nonlinear Energy Storage Model}
Because reactive power cannot be stored, $\Sstor = \Pstor+ \imag 0$, and the amount of power stored is separated into charge $\Pcstor$ and discharge $\Pdstor$ power values, both of which are positive,
\begin{IEEEeqnarray}{C}
\Pstor = \Pdstor - \Pcstor  \label{eq:ac-6}
     \end{IEEEeqnarray}
 Charging and discharging are typically complementary,
\begin{IEEEeqnarray}{C}
\Pdstor \cdot \Pcstor = 0 \label{eq:ac-7} \label{eq_complementarity}
     \end{IEEEeqnarray}
The energy content in the buffer changes when charging and discharging, which also incurs losses which are captured by $\etacstor$ and $\etadstor$ as follows,
\begin{IEEEeqnarray}{C}
\forall \indextime \in \setTime \setminus 1: \Ect - \Ectprev = \Tt \left(\etacstor \Pcstor - \frac{\Pdstor}{\etadstor} \right) \label{eq:ac-8}
     \end{IEEEeqnarray}
when using the end-point time discretization. Special care is taken to initialize the energy buffer in a certain state at the first time step,
\begin{IEEEeqnarray}{C}
\indextime=1: \Ect - \Ectinit = \Tt \left(\etacstor \Pcstor - \frac{\Pdstor}{\etadstor} \right)
\end{IEEEeqnarray}
It is possible to avoid defining an initial energy content by adding the a constraint that the final energy content must be higher than the initial one. For instance,
\begin{IEEEeqnarray}{C}
\Ectend \geq \Ectone.
\end{IEEEeqnarray}
If so included, one can  enforce a specific end-time value instead.
\subsection{Variable Bounds }
  We define  the total power rating of the converter  as the sum of the per-phase values.
    \begin{IEEEeqnarray}{C}
\Scrating = \sum_p \Scratingp.
  \end{IEEEeqnarray}
  If per-phase values are unknown, we can derive them from the total through $\Scratingp = \Scrating / |\setconductor|$.
The magnitude bounds on the complex variables are,
\begin{IEEEeqnarray}{C}
 (\Pcpt)^2 + (\Qcpt)^2 \leq  |\Uipt|^2 (\statusct \cdot \Icratingp)^2, \label{eq_current_bounds}\\
 (\Pcpt)^2 + (\Qcpt)^2 \leq (\statusct \cdot \Scratingp)^2. \label{eq_power_bounds}
\end{IEEEeqnarray}
It is not obvious whether just the current rating \eqref{eq_current_bounds} or power ratings \eqref{eq_power_bounds} are to be considered, so we include both. Note that both ratings are defined per-phase, thereby enabling two/single-phase connected systems as edge cases through conductor-wise values set to 0.
In rectangular variables the converter bounds are,
\begin{IEEEeqnarray}{C}
-\statusct \cdot\Scratingp \leq \Pcpt \leq \statusct \cdot\Scratingp,\\
-\statusct \cdot\Scratingp \leq \Qcpt \leq \statusct \cdot\Scratingp,\\
- \statusct  \cdot \Scrating \leq \Qint \leq \statusct  \cdot \Scrating,\\
 -\Scrating \leq \Pstor \leq \Scrating. \label{eq:ac-12}
\end{IEEEeqnarray}
The storage subsystem bounds are,
\begin{IEEEeqnarray}{C}
0 \leq \Pcstor \leq \Pcstorrating, \\
 0 \leq \Pdstor \leq \Pdstorrating, \\
0 \leq \Ect \leq \Erating, \\
0 \leq \Uiptmin \leq |\Uipt| \leq \Uiptmax.  \label{eq:ac-18}
\end{IEEEeqnarray}

Together, this system of equations and inequalities defines the most general form of the storage model proposed by this work. Note that model the only nonconvex constraint is \eqref{eq_complementarity}.

\subsection{Data Model}

\begin{table}[tb]
\setlength{\tabcolsep}{3pt}
  \centering
   \caption{Parameters}\label{tab_params}
    \begin{tabular}{l l l l l  }
\hline
Symbol & SI& data model  & condition & meaning \\
\hline
$\Tt$ &  s&  h & $>0$  &  time step length   \\
\hline
$\Uiptmin$ &  V&  pu & $\geq0$  & voltage lower bound  \\
$\Uiptmax$ &  V&  pu & $\geq \Uiptmin$  &  voltage upper bound \\
\hline
$\statusct $&  - &  - & $\in \{0,1\}$  & converter status  \\
$\Sext $& VA& MVA & $-$  & external flow  \\
\hline
$\Scrating, \Scratingp$ &  VA&  MVA & $\geq0$  &  apparent power rating \\
$\Icrating, \Icratingp $&  A &  MA & $\geq0$  & current rating  \\
$\etadstor $&  -&  - & $>0, \leq1$  &  discharge efficiency \\
$\etacstor$ &  -&  - & $\geq0, \leq1$  & charge efficiency  \\
$\Ectinit $&  J&  MWh & $\geq0$  & initial energy content  \\
$\Erating $&  J&  MWh & $\geq0$  & energy rating  \\
$\Pcstorrating $&  W&  MW & $\geq0$  & charge power rating  \\
$\Pdstorrating$ &  W&  MW & $\geq0$  & discharge power rating  \\
$\Zcp$ &  $\Omega$ & pu & $- $  & converter impedance \\
\hline
\end{tabular}
\end{table}

Table \ref{tab_params} summarizes the constant parameters defined in the mathematical models.
We note that while all equations in this report are given in SI units, the user-facing data model utilizes the indicated engineering units.
Table \ref{tab_examples} illustrates how the data model can be parameterized to capture common storage technologies such batteries and pumped hydro can be achieved.  Time step length is uniform at $\Tt$ = 0.25h and number of time steps $|\setTime|= 96$.

\begin{table}[tbh]
  \centering
   \caption{Illustrative Parameterization for Specific Technologies}\label{tab_examples}
    \begin{tabular}{l l l l l  }
\hline
Parameter & Unit  & BESS & PHS & Flywheel \\
\hline
$\statusct $&   - & 1  & 1  & 1 \\
$\Sext $&  MVA & standby  & flow into & standby \\
&   &  power &  upper reservoir & power \\
&  MVA & 0.00005 & time series & 0.003 \\
\hline
$\Scrating$ &  MVA & 0.005   & 1000  & 0.1 \\
$\etadstor $&  - & 0.95  & 0.90 & 0.92 \\
$\etacstor$ &  - & 0.95 & 0.90 & 0.92 \\
$\Ectinit $&  MWh & 0.010  & 1000 & 0.015 \\
$\Erating $&  MWh & 0.010  & 2000 & 0.030\\
$\Pcstorrating $&  MW & 0.005  & 1.0  &  0.1\\
$\Pdstorrating$ &  MW & 0.005 & 0.72 &  0.1\\
$\Rcp$ &pu  &0.1   & 0 & 0 \\
$\Xcp$ &pu  & 0   & 0.1 & 0\\
\hline
\end{tabular}
\end{table}

\section{Reformulation, Relaxation and Approximation}
\label{sec_model_relax}
The key step to integrate the storage models into OPF, is to aggregate the storage systems for a specific bus, and then add the aggregated variables $\Pcpt, \Qcpt$ to the nodal power balance equations for each conductor.

Nevertheless, this approach  is not immediately applicable to power system optimization tasks that do not typically consider multi-conductor network models, such as balanced OPF and production cost models.
Furthermore, such a  nonlinear model can present significant computational challenges in large-scale optimization contexts.
To make this model more widely applicable, this section develops a series of simplifications for using the proposed model in convex and approximate power flow models.

\paragraph*{Balanced Power Flow Approximation}
The most natural approximation of the proposed model is a single-phase balanced power flow approximation.  This is easily accomplished
by considering the single-phase case of \eqref{eq_balance_loss} in the general model, namely,
\begin{IEEEeqnarray}{C}
    \Sct + \Pstor =  \imag \Qint + \Sext+  \Zc |\Ict|^2  \label{eq:ac-5}
\end{IEEEeqnarray}
A similar transformation can be applied to any of the model variants presented in this section.

\paragraph*{Discrete Complementarity Formulation}
The charge-discharge complementarity constraint \eqref{eq:ac-7} is a non-convex nonlinear constraint that has undesirable numerical properties.  Consequently, it is common to reformulate this constraint by introducing a binary variable,
\begin{IEEEeqnarray}{C}
 \bc \in \{0,1\} \label{eq_integrality}
 \end{IEEEeqnarray}
that indicates which of the charge/discharge variables is active at a given point in time by replacing \eqref{eq_complementarity} with,
\begin{IEEEeqnarray}{C}
 0 \leq \Pcstor \leq \Pcstorrating \cdot  \bc \label{eq_complementarity_mi_1}\\
 0 \leq \Pdstor \leq \Pdstorrating \cdot(1-  \bc)  \label{eq_complementarity_mi_2}
\end{IEEEeqnarray}
The discrete nature of this formulation does not typically present a overwhelming burden in commercial optimization software.  However, if such a problem is encountered, the integrality of this indicator variable can be relaxed as follows, providing a fast continuous relaxation of the complementarity requirement,
\begin{IEEEeqnarray}{C}
0 \leq \bc \leq 1.
\end{IEEEeqnarray}

\paragraph*{Convex-Quadratic Relaxation}
Convex quadratic relaxations of the power flow equations have gained interest recently and we observe that the proposed storage model has a natural relaxation in terms of the squared magnitudes of voltage and current.  First, the voltage and current squared expressions are lifted to new variables as follows:
\begin{IEEEeqnarray}{C}
  |\Uipt|^2    \rightarrow \Wipt, \\
    |\Icpt|^2 \rightarrow \Lcpt .
\end{IEEEeqnarray}
Using these lifted variables one can rewrite \eqref{eq_complex_power_def} and \eqref{eq_balance_loss} as convex constraints,
\begin{IEEEeqnarray}{C}
     (\Pcpt)^2 + (\Qcpt)^2 \leq \Wipt \Lcpt, \label{eq_lifted_power_relax}\\
       \sum_{\indexconductor \in \setconductor}  \Scpt + \Pstor = \imag \Qint + \Sext + \sum_{\indexconductor \in \setconductor} \Zcp \Lcpt. \label{eq_lifted_storage_relax}
\end{IEEEeqnarray}
Note that \eqref{eq_lifted_power_relax} is a rotated second order cone, and therefore convex.
It is advantageous to combine this relaxation with the discrete complementary formulation resulting in a mixed-integer second-order conic optimization problem, which can be solved with highly optimized commercial software packages.

\paragraph*{Linearized Approximation}
Given the high performance of the DC power flow approximation, a linear active-power-only approximation of the storage model is proposed by replacing the converter balance \eqref{eq_balance_loss} with,
\begin{IEEEeqnarray}{C}
  \sum_{\indexconductor \in \setconductor}  \Pcpt + \Pstor = \Pext \label{eq_storage_approx}.
\end{IEEEeqnarray}
Similar to the DC power flow approximation, this formulation does not capture the copper losses incurred from operating the converter, but it does benefit from a significant reduction in model size because reactive power is ignored.

\section{Computational Validation} \label{sec_study}

The generic storage model proposed in the previous sections is implemented in the open-source power network optimization software PowerModels v0.17 \cite{Coffrin2017}.  This section presents two proof-of-concept validation studies demonstrating the soundness of the model and how the dispatch of a storage device can vary across different formulations of the power flow equations.  Throughout this section the following solvers were used: Ipopt v12.4 \cite{Wachter2006} for the NLP version of AC model; Juniper v0.6 \cite{Kroger} using Ipopt as an NLP sub-solver for the MINLP version of AC model; Gurobi v9.0 \cite{gurobi} for the MIP and MISOCP models presented by the DC and SOC models.

\subsection{Test Case}
A multi-period AC-OPF test case is required to study the storage model proposed in this work.  In the interest of utilizing open-access data, a test case was constructed by combining the 14-bus network available in the PGLib-OPF v19.05 \cite{Babaeinejadsarookolaee2019}  with the hourly load profile scalars provided in the RTS 96 test case for a typical summer weekday.  Linear interpolation was used to increase the load profile's resolution from hourly to 15 minute increments (\textit{i.e.}, $\Tt$ = 0.25h, $|\setTime|= 96$).  A moderately sized storage device was added to Bus 13 in the test case with the following parameters, $\Ectinit = 1, \Erating = 200, \Pcstorrating = 100, \Pdstorrating = 75, \etacstor = 0.85, \etadstor = 0.90,  Z = 0.1 + \imag 0.01, \Scrating = 1000$; rather than focus on a specific storage technology, this invented device was designed to exercise the key mathematical features of the proposed model.  Finally, a quadratic cost term of $0.2$ \$/MWh$^2$ was added to the generators to provide a cost incentive to charge the storage during off-peak times.

The proposed extended version of the 14-bus network is suitable for single-phase multi-period AC-OPF studies.  To test the multi-phase variant of the proposed storage model a three-phase extension of this network was developed by creating three replicates of network (one for each phase) and splitting the load by 36\%, 33\%, and 31\% on phases A, B, and C respectively.  This proposed network is a highly stylized version of a multi-phase network model, though it is sufficient to highlight some of the non-trivial interactions that can occur between multi-phase storage devices and generator cost functions.

\subsection{AC Optimal Power Flow}
The first and simplest experiment considers solving the multi-period AC-OPF with storage; note that both models are only guaranteed to converge to a locally optimal solution. Both the NLP and MINLP variants of the AC-OPF with storage model were considered (see Table \ref{tbl:obj_rt}).  It was observed that both formulations found similar quality solutions.  While the NLP variant was significantly faster to converge it also occasionally suffered from numerical issues, suggesting the MINLP formulation is preferable for robustness.

\begin{figure}[t]
    \begin{center}
    \includegraphics[width=8.8cm]{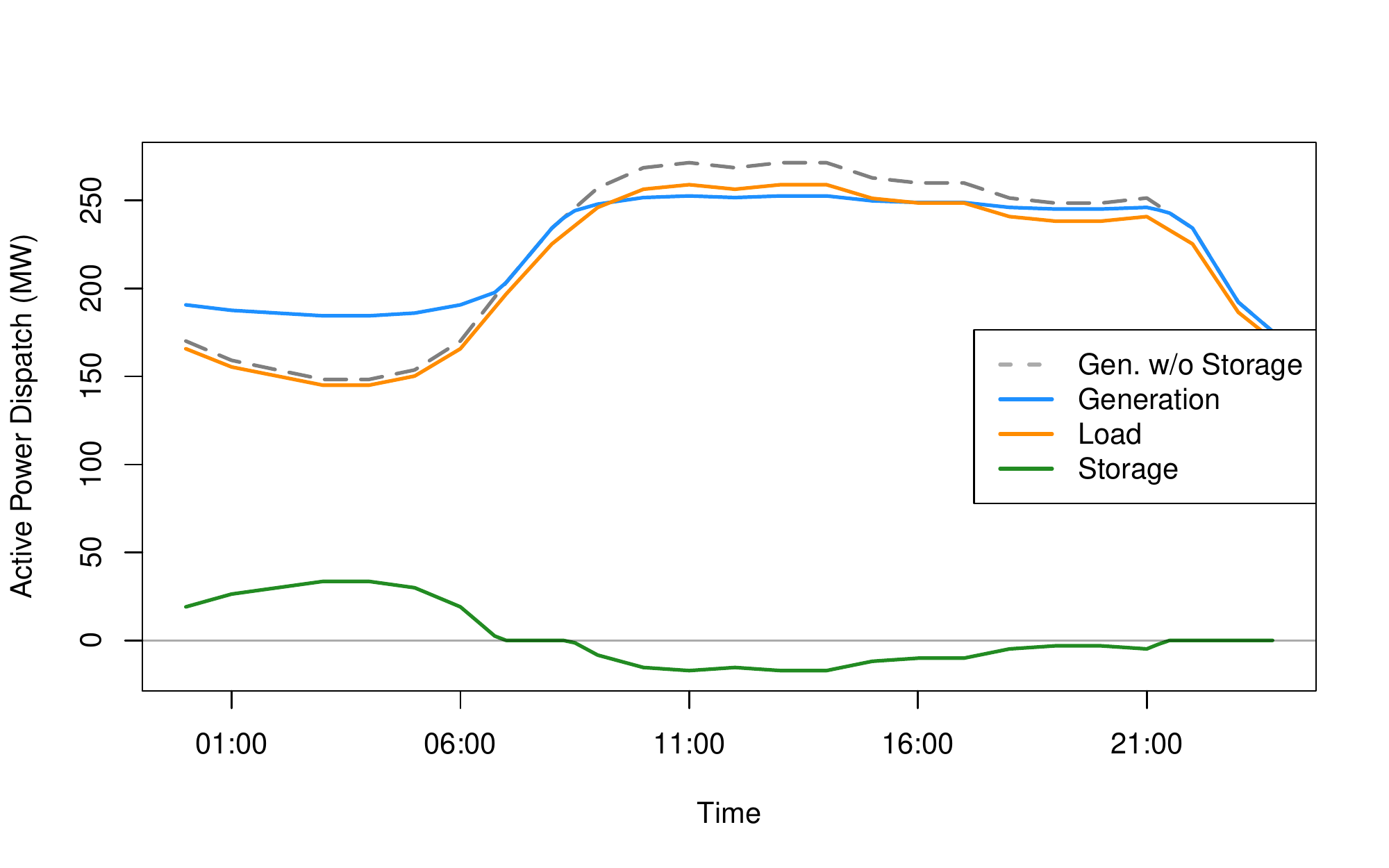}
    \end{center}
    \vspace{-0.6cm}
    \caption{A comparison of storage impacts to generator dispatch in single-phase AC power flow. Positive/negative storage dispatch indicates charging/discharging, respectively.}
    \vspace{-0.2cm}
    \label{fig:ac_storage}
\end{figure}

The operation profiles presented in Figure \ref{fig:ac_storage} are qualitatively what is expected from the introduction of a storage device.  A time shift of the load allows the total generation cost to be reduced from 882,439 to 871,971.  This is accomplished by storing energy in the evening that is expended during the day, as illustrated by the peak-shaving behavior in Figure \ref{fig:ac_storage}.  However encouraging these results are, it is important to verify the quality of the proposed charge/discharge schedule due to the non-convex nature of these models.  The SOC relaxation and DC approximations are natural choices for verification because their convexity ensures globally optimal schedules.

\subsection{Power Flow Model Comparison}
\begin{figure}[t]
    \begin{center}
    \includegraphics[width=8.8cm]{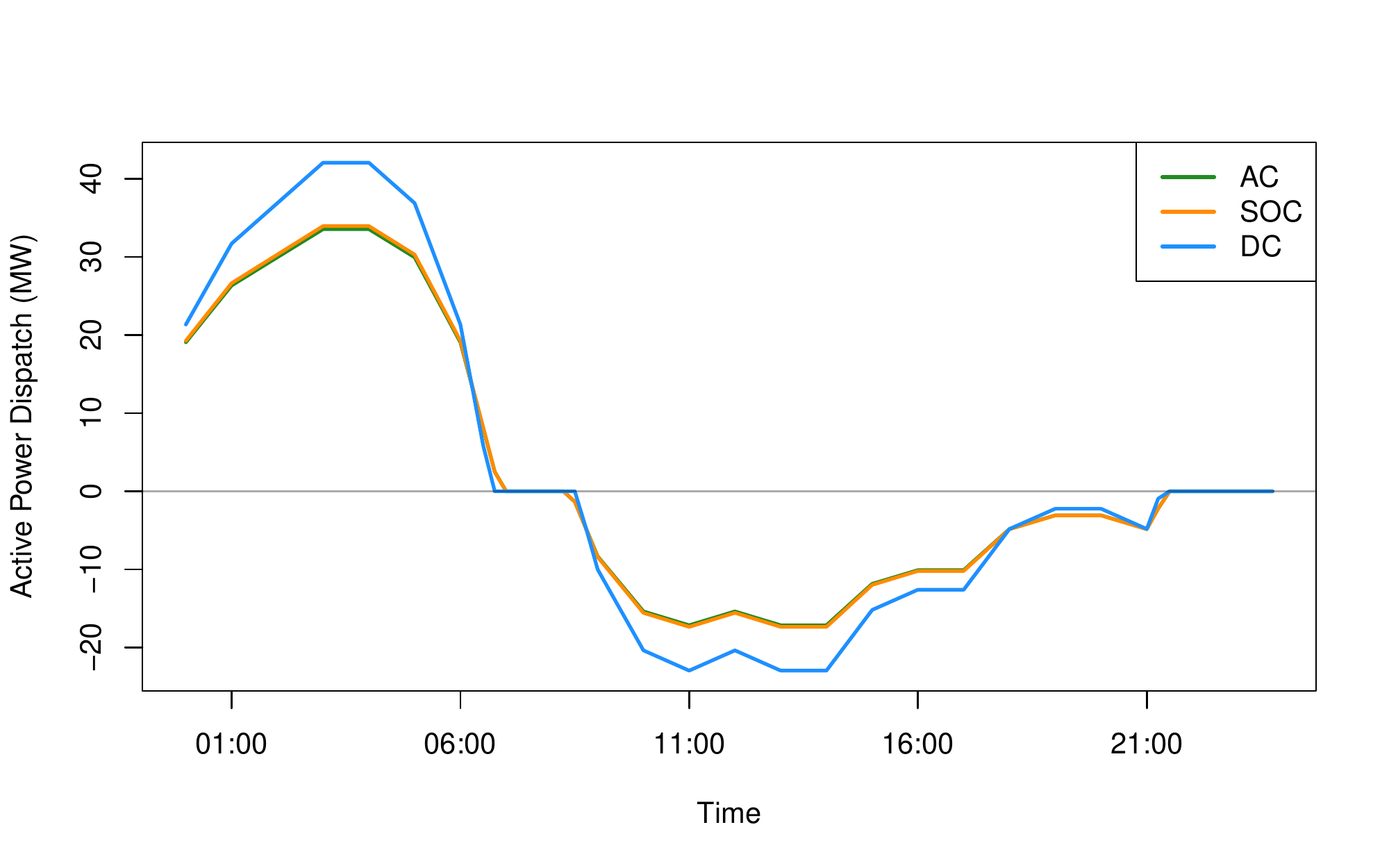}
    \end{center}
    \vspace{-0.6cm}
    \caption{A comparison of power flow models on storage dispatch. Positive/negative storage dispatch indicates charging/discharging, respectively.}
    \label{fig:ac_storage2}
\end{figure}

The second experiment compares the charge/discharge schedule proposed by three power flow models: the non-convex AC power flow, the convex nonlinear SOC power flow and the linear `DC' power flow.  The results of this experiment are highlighted in Table \ref{tbl:obj_rt} and Figure \ref{fig:ac_storage2}.  Focusing on Figure \ref{fig:ac_storage2}, the first observation is that AC and SOC dispatch schedules are remarkably similar, which suggests that the AC solution is of high quality.  Furthermore, because the SOC model is a convex relaxation and provides a lower bound on the optimal solution, we can compute an optimally gap of less than $0.17\%$ on the AC solution, based on the values reported in Table \ref{tbl:obj_rt}.  The second observation in Figure \ref{fig:ac_storage2} is that the DC power flow generally tracks the behavior of the other two models but is notably more aggressive in the charge/discharge schedule.  A preliminary sensitivity investigation suggested that this discrepancy can be explained in large part by the lack of converter loss modeling in this linear approximation.

\begin{table}[tbh]
  \centering
   \caption{Storage Model Cost and Runtime Comparison}\label{tbl:obj_rt}
    \begin{tabular}{r r r r r  }
\hline
Model & Storage Equations & Objective & Runtime (s)  \\
\hline
AC Base & none & 882\,439 & 5.06 \\
AC-NL & \eqref{eq:ac-5},\eqref{eq_complementarity} & 871\,971 & 82.26 \\
AC-MI & \eqref{eq:ac-5},\eqref{eq_complementarity_mi_1},\eqref{eq_complementarity_mi_2} & 871\,971 & 686.48 \\
SOC-MI & \eqref{eq_lifted_power_relax},\eqref{eq_lifted_storage_relax},\eqref{eq_complementarity_mi_1},\eqref{eq_complementarity_mi_2} & 870\,519 & 5.43 \\
DC-MI & \eqref{eq_storage_approx},\eqref{eq_complementarity_mi_1},\eqref{eq_complementarity_mi_2} & 807\,625 & 0.12 \\
\hline
\end{tabular}
\end{table}

\subsection{Single-Phase vs Multi-Phase}
\begin{figure}[t]
    \begin{center}
    \includegraphics[width=8.8cm]{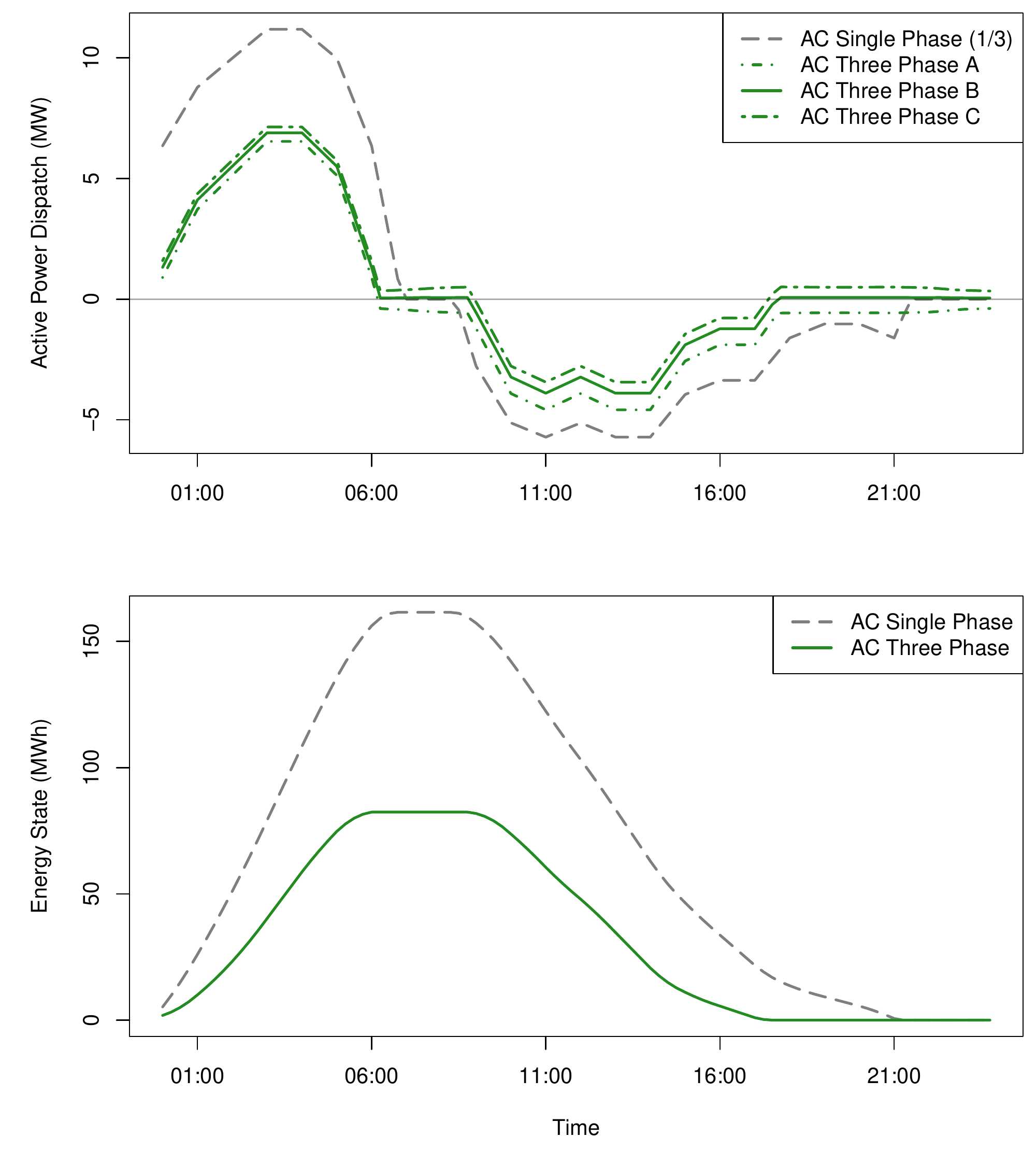}
    \end{center}
    \vspace{-0.6cm}
    \caption{A comparison of storage dispatch (top) and energy state (bottom) in single-phase and multi-phase power flow models.}
    \vspace{-0.5cm}
    \label{fig:ac_3p_storage}
\end{figure}

This third experiment highlights the value of the proposed model in a simplified multi-phase setting.  Figure \ref{fig:ac_3p_storage} provides a comparison of the storage dispatch and state in both the AC single-phase and AC three-phase settings.  The first observation is that the total utilization of the storage device is reduced significantly.  This is driven primarily by the added value of the converter to exchange power across phases, which reduces the relative value of storing energy over time.
This effect is highlighted by the storage dispatch in the last few hours of the day where energy is consumed on Phases B/C and delivered to Phase A to reduce the cost of generating power on the more heavily loaded phase.  This stylized example highlights the importance of considering converter models and phase unbalance information when evaluating the sizing, operation and value of storage devices.

\section{Conclusions}
\label{sec_conclusions}
Motivated by the increasing role of storage in future energy systems, this work proposed a generic and flexible storage model that can be leveraged in a variety of power system optimization studies, ranging from operation scheduling to production cost modeling.  Based on a computational validation of the proposed model, we conclude that the details of both the storage device and its converter can have a non-trivial impact on the utilization and value of energy storage.  These results highlight the value of the proposed model in switching between different levels of modeling fidelity.

The proposed model has been implemented in PowerModels since v0.9 \cite{Coffrin2017} and PowerModelsDistribution since v0.2 to provide easy access by the research community.  A core feature of the PowerModels framework is that it allows the user to easily switch between different types of power flow formulations (\textit{e.g.}, non-convex, convex and linear), which provided a clear path to implement the storage model variants proposed in this work.  In future work we will continue testing the proposed model in more complex optimization contexts, such as OPF with unit commitment and optimal storage sizing, to better understand how storage model fidelity impacts optimal decisions.  We also plan to continue validation efforts with direct comparisons to the detailed storage models provided by power system simulators.

\section{Acknowledgements}
This work was supported by funding from the U.S. Department of Energy’s (DOE) Office of Electricity (OE) as part of the CleanStart-DERMS project of the Grid Modernization Laboratory Consortium, and by the U.S. Department of Energy through the Los Alamos National Laboratory LDRD Program and the Center for Nonlinear Studies.

\bibliographystyle{IEEEtran}

\end{document}